\documentclass[aps,prb,twocolumn,floatfix,showpacs]{revtex4-1}
\usepackage{graphicx,bm}
\usepackage{times}
\usepackage{amsmath,amssymb}
\usepackage{epstopdf}

\begin{document}
\title{Optical absorption spectroscopy in hybrid systems of plasmons and excitons}

\author{Chen-Yen Lai}
\author{S. A. Trugman}
\author{Jian-Xin Zhu}
\affiliation{Theoretical Division, Los Alamos National Laboratory, Los Alamos, New Mexico 87545, USA}
\affiliation{Center for Integrated Nanotechnologies, Los Alamos National Laboratory, Los Alamos, New Mexico 87545, USA}
\date{\today}


\begin{abstract}
  Understanding the physics of light emitter in quantum nanostructure regarding scalability, geometry, structure of the system and coupling between different degrees of freedom is important as one can improve the design and further provide controls of quantum devices rigorously.
  The couplings between these degrees of freedom, in general, depends on the external field, the geometry of nano particles, and the experimental design.
  An effective model is proposed to describe the plasmon-exciton hybrid systems and its optical absorption spectrums are studied in details by exact diagonalization.
  Two different designs are discussed: nano particle planet surrounded by quantum dot satellites and quantum dot planet surrounded by nano particle satellites.
  In both setups, details of many quantum dots and nano particles are studied, and the spectrums are discussed in details regarding the energy of transition peaks and the weight distribution of allowed transition peaks.
  Also, different polarization of external fields are considered which results in anisotropic couplings, and the absorption spectrums clearly reveal the difference qualitatively.
  Finally, the system will undergo a phase transition in the presence of attractive interaction between excitons.
  Our work sheds the light on the design of nano scale quantum systems to achieve photon emitter/resonator theory in the plasmon-exciton hybrid systems.
\end{abstract}

\maketitle

\section*{Introduction}
The rapidly growing field of research on light-matter interactions has opened an arena to manipulate light through nanoscale quantum devices~\cite{Tame:2013iv}.
Using metallic nanoparticles or nanostructures can increase the local density of electromagnetic modes and enable spontaneous emission control of optical transitions~\cite{Farahani:2005kn}.
On the metal surface, there is a collective, wave-like motion of electrons which is the most fundamental element in quantum plasmonics and is known as surface plasmons~\cite{Maier:2007wq,Tame:2013iv,Chang:2007fk,Madsen:2017bs,Miyata:2015}.
Moreover, due to dissipation, these plasmon modes decay quickly with a relaxation time scale around $10\!-\!100fs$ which can be a useful tool in ultrafast signal processing, but is not ideal for effective resonators.
Several designed devices, including cavity quantum electrodynamics~\cite{Rice:1996iz}, conjugated polymer molecules~\cite{Wu:2008ib}, emitter with nanowires~\cite{Chang:2006gx,Akimov:2007ex,Saunders:2006bf,Singh:2011ey,Ming:2009de}, emitter with nano particles~\cite{Naiki:2013il,Pan:2013eo,Ji:2015bz,Hollingsworth:2015ir}, are realized in experiments in order to manipulate light and provide different functionalities, such as single-photon sources, transistors, and ultra-compact circuitry at the nanoscale.

To understand the physics and effects on intrinsic energy scales and structure of the systems, it is important to consider the quantum nature, such as collective excitations or particle-hole pairs~\cite{Tame:2008hm}, as well as different forms of the coupling between these degrees of freedom.
To fully characterize the effects and features of these processes,
one needs to properly model the system in a quantum mechanical manner and solve it in a quantum mechanical framework.
Also, the decoherence and dissipation effects play a crucial role in these systems as well, as some studies using a master equation~\cite{Artuso:2010ed,Ridolfo:2010ed,Hayati:2016ep} have also reported.
On the other hand, the plasmon-assisted resonance energy transfer~\cite{Pustovit:2011hs,Pustovit:2013bc,Pustovit:2014gm} and cooperative emission from the Dicke mechanism~\cite{Gross:1982js,BRANDES:2005gd} are also discussed in the literature for both superradiance and subradiance~\cite{Pustovit:2010cq}.
Although the plasmon-exciton hybrid system was proposed and studied~\cite{Manjavacas:2011jr,Chen:2015kr,Zhang:2011ie,Shahbazyan:2017ih,Shahbazyan:2017ce} in various ways where the system with only a few metallic NPs and one QD is considered and solved in a certain approximation, some important issues, such as the scalability of the system, the effects from coupling and interaction between plasmons and between excitons, and modeling the coupling between these two degrees of freedom are still needed to be addressed, solved in quantum mechanical fashion and eventually make connections to the experiments.

In this article, we adopted an effective model to describe the plasmon-exciton hybrid system where the plasmons and excitons are coupled to external electric probing fields and the dominant dipole-dipole interactions between plasmon and exciton.
The model is solved in exact diagonalization~\cite{Alfonsi:2010jg,Zhang:2010bh} which captures the features of the optical absorption spectrum without approximations or prior assumptions.
The advantage of the method is that we can consider system sizes achieved in the experiments~\cite{Hollingsworth:2015ir}, include all relevant degrees of freedom and coupling forms in the model, and determine the optical absorption spectrum without bias.
The hybrid systems have large metallic nano particles (NP) which can be properly described by plasmons~\cite{Naiki:2013il,Pan:2013eo,Ji:2015bz} and small size quantum dots (QD) which have excitonic degrees of freedom.
Two different structures of NPs and QDs are considered in this article.
One is to put several QDs around a NP (nano particle planet surrounded by quantum dot satellites), and another is to have a QD in the center surrounded by NPs (quantum dot planet surrounded by nano particles satellites).
The former is proposed for emission enhancement~\cite{Naiki:2013il}, and the later to be a potential platform for a nanoantenna~\cite{Ji:2015bz,Savasta:2010gb,Marinica:2013ip} and quantum information devices~\cite{Pelton:2010fi}.

\section*{Plasmon-Exciton hybrid system}\label{sec:model}
The optical absorption spectrum (OAS) can be obtained from Fermi's Golden rule in the linear response regime,~\cite{Manjavacas:2011jr} which is given by
\begin{eqnarray}\label{eq:fgr}
  \sigma(\omega)&=&\sum_f\vert A_f\vert^2\delta(E_f-E_i-\omega),
\end{eqnarray}
with $A_f=\langle f;n-1\vert\mathcal{H}^\prime\vert i;n\rangle$ and $\hbar\!=\!1$.
Here $\vert i(f)\rangle$ represents the initial (final) state of the system; $n$ is the number of initial photons with energy $\hbar\omega$; and $\mathcal{H}^\prime$ is the Hamiltonian coupling the system to the external photon field which has the form $\mathcal{H}^\prime\!\propto\!Aa^\dagger\!+\!A^\dagger a$ in the rotating wave approximation.
Here $a$ and $A$ are annihilation operators for external photon and single excitation of the system respectively.
Therefore, $A$ connects the initial and final state of the system.
Eq.~\eqref{eq:fgr} can be reexpressed by replacing the delta function $\delta(x)\!=\!\frac{1}{\pi}\lim_{\eta\rightarrow0^+}\text{Im}\left(\frac{1}{x-i\eta}\right)$ to arrive at
\begin{eqnarray}
  \sigma(\omega) &\propto& \frac{1}{\pi}\lim_{\eta\rightarrow0^+}\text{Im}\left\{
  \sum_f\frac{\langle i \vert A \vert f \rangle\langle f \vert A^\dagger \vert i\rangle}{E_f - E_i - \hbar\omega -i\eta}
  \right\}\label{eq:oas}
\end{eqnarray}
where $\vert i \rangle$ is the initial ground state $\vert 0 \rangle$.
$\eta$ is a small number to ensure the convergence which is also related to the dephasing time of the NP or QD.

The systems considered in this article consist of both QDs and NPs.
In general, both can be excited by an external light source and larger particles interact more strongly than smaller particles due to the larger absorption cross section~\cite{Manjavacas:2011jr}.
To avoid confusion, we note that the calculated adsorption on NP takes into account all coherent processes in the hybrid system.
For most of the calculation, we consider metallic particles interact more strongly with an external electric field than dielectric particles (off resonance) and only absorption signals from NPs are presented in the first part of this article.
The effects of nonzero simultaneous absorption from both NPs and QDs are discussed later in this article (Fig.~\ref{fig:Ratio} and accompanying subsection).
The effects of different absorption rate from both NPs and QDs are discussed later.
In experiments, a metallic NP can be synthesized from, for instance, metal nanocrystals gold or silver with semiconductor quantum-shell~\cite{Ji:2015bz,Naiki:2013il,Hollingsworth:2015ir} and can be properly described by plasmons~\cite{Manjavacas:2011jr,Tame:2013iv} which can be polarized and excited by an external electric field and have moments.
For simplicity, we consider that the NP only has active dipole moments and the higher order multipolar modes~\cite{note2} are ignored for most of the calculation.
The effects from quadrupole moment of NP are briefly discussed in the Methods section.
As the size of the NP becomes larger, multipolar modes will be excited and actively coupled to excitons and other plasmon modes~\cite{Ina:2011,Wing:2016,Giannini:2010pp,Pustovit:2010cq}.
On the other hand, the semiconductor QDs can be engineered, like CdSe/CdS/ZnS~\cite{Naiki:2013il,Ji:2015bz,Karan:2015ft,Hollingsworth:2015ir},
and they can be modeled as two-level systems for the exciton dynamics~\cite{gammon2002,Manjavacas:2011jr,Savasta:2010gb,Artuso:2010ed,salomon2012a,lopata2009}.
We model the system by the following Hamiltonian,
\begin{eqnarray}
  H_{\text{NP}} &=& \sum_{i,m}E_dd_{im}^\dagger d_{im} 
  - \sum_{mn}\sum_{\langle ij\rangle}J^{(ij)}_{d,mn}(d_{im}^\dagger d_{jn}+h.c.),\\
  H_{\text{QD}} &=& \sum_{i,m}E_cc_{im}^\dagger c_{im} 
  - \sum_{mn}\sum_{\langle ij\rangle}J^{(ij)}_{c,mn} (c_{im}^\dagger c_{jn}+h.c.).\label{eq:Hqd}
\end{eqnarray}
Here, $c_{im}^\dagger(c_{im})$ is the creation (annihilation) operator for excitons on the QD-$i$ with $m$-mode and $d^\dagger_{jm}(d_{jm})$ is the creation (annihilation) operator for plasmons on NP-$j$ with mode $m(=x,y,z)$ of angular momentum $l\!=\!1$ state.
The $\langle ij\rangle$ indicates nearest neighbors sites.
The internal energy levels are related to the physical size and material of the NP and QD.
The excitons are treated as hard-core bosons and the plasmon operators obey the Bose-Einstein statistics.
The coupling between pairs of plasmons (excitons) are in general anisotropic and are determined from
the interaction energy of their dipole operators.
The excitons and plasmons also interact with each other and the corresponding coupling is given by~\cite{Nordlander:1986hx,Nordlander:2004bh}
\begin{eqnarray}
  H_{\text{dc}}=-\sum_{m,n}\sum_{\langle ij\rangle}\Delta^{(ij)}_{dc,mn}(d^\dagger_{im}c_{jn}+h.c.) \;.
\label{eq:NP-QDcoupling}
\end{eqnarray}
This is most important coupling between NP-QD pairs in the linear response regime. It depends on their intrinsic properties, and the distance and angle between the pairs.
In general, when both the QD and NP are coupled to the external probing photon source, the transition operator in Eq.~\eqref{eq:oas} is $A\!=\!C_{NP}\sum_id_{im}+C_{QD}\sum_ic_{im}$, which excites a state with zero momentum of the $m$-mode.
Due to the larger absorption cross section, we will first focus on the cases where $C_{NP}\!\gg\!C_{QD}$. The situations with varying absorption ratio will be discussed later.
Our methods also can be applied to excitation with non-zero momentum states.
Besides that, the long range coupling between different degrees of freedom can also be included in our formula.
The OAS from Eq.~\eqref{eq:oas} along with the system Hamiltonian $H\!=\!H_{\text{NP}}\!+\!H_{\text{QD}}\!+\!H_{\text{dc}}$ can be solved by exact diagonalization if the system is small or by Lanczos methods~\cite{Zhang:2010bh,Avella:2013jm,Deng:2016bl,Lai:2016kh} for larger systems.
A brief summary of the Lanczos method is provided in the Methods section.
Exact diagonalization gives the full spectrum of Hamiltonian, and all transitions are captured.
If the second approach is chosen, the number of peaks in OAS is limited by the number of Lanczos vectors used.
We ensure that major transitions are captured by using $A\vert i\rangle$ as the first Lanczos vector and iterate until the norm of next Lanczos vector is smaller than $10^{-12}$.
The parameter $\eta$ will be set to $5meV$ throughout this article and can be taken as the lifetime of the excitations, which will broaden the transition peak.

\section*{Results and discussion}\label{sec:res}
Two different structures are considered in this article: NP-planet-QD-satellite illustrated in Fig.~\ref{fig:6QDsJcW}a and QD-planet-NP-satellite in Fig.~\ref{fig:6NPsJdW}a.
Both structures are constrained because
the physical size of a NP is much larger than a QD in the experiments~\cite{Naiki:2013il,Pan:2013eo,Ji:2015bz,Karan:2015ft}.

\subsection*{Isotropic coupling}\label{sec:iso}
Assuming the external probing field is perpendicular to the plane where QDs and NPs are assembled, we can consider only a single mode ($m\!=\!z$) on each NP and QD and all couplings $J^{<ij>}_{d,mn}\!=\!J_d$, $J^{<ij>}_{c,mn}\!=\!J_c$, and $\Delta_{dc,mn}^{(i,j)}\!=\!\Delta_{dc}$ are isotropic for simplicity.

\begin{figure}[h]
  \begin{center}
    \includegraphics[width=0.48\textwidth]{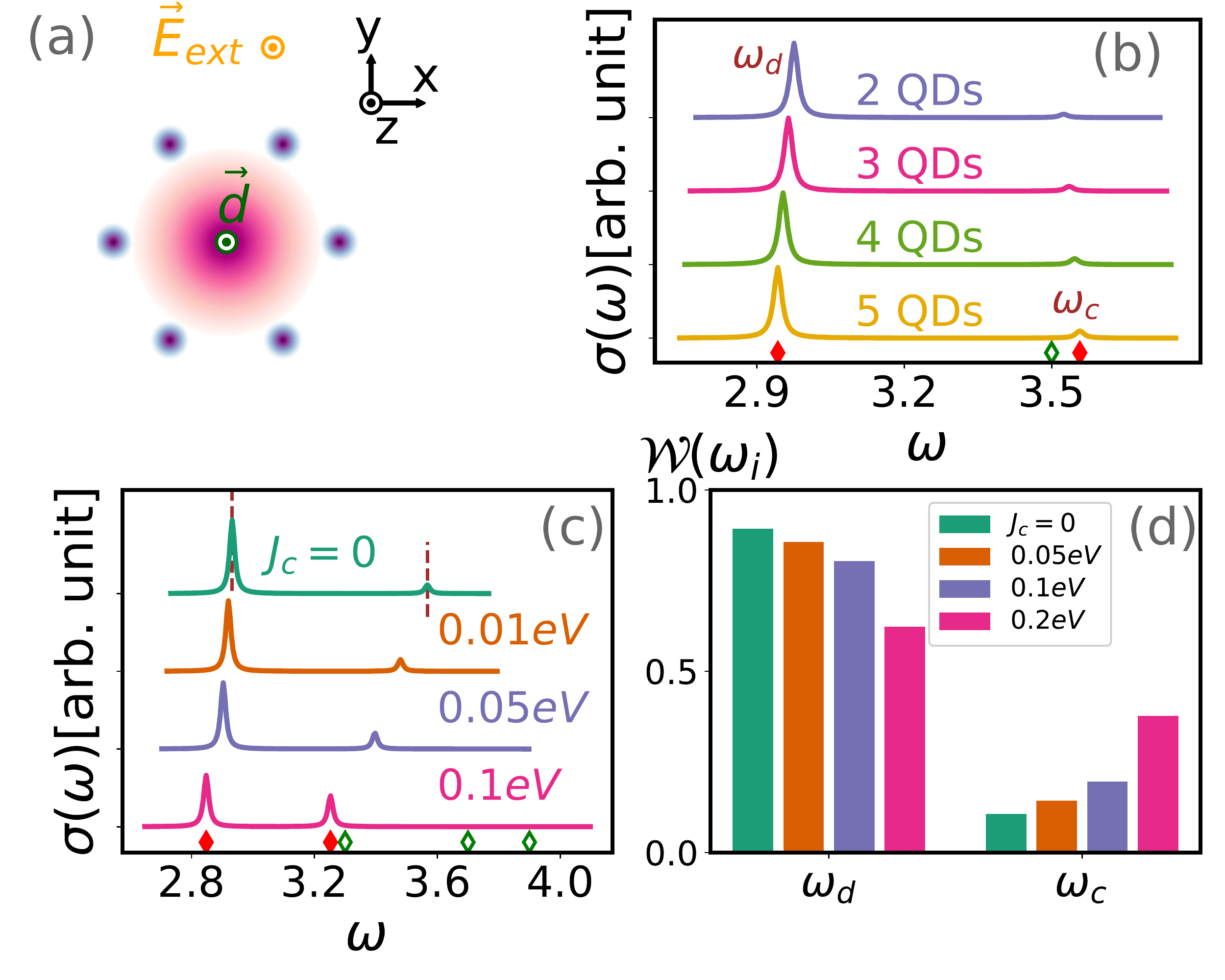}
    \caption{
      (a) The system has one NP planet and six QD satellites surround it in the $x$-$y$ plane.
      The probing field is along the $z$-axis.
      (b) The OAS of systems with one NP planet coupled to different numbers of QD satellites.
      The coupling between QDs are set to $J_c\!=\!0$.
      Both transition peaks $\omega_{d,c}$ shift with different numbers of QDs in the system, as in Eq.~\eqref{eq:omegadc}.
      The empty (filled) diamond symbols under the curve mark the transition peaks with (non-) zero weight.  There are degenerate eigenvalues on top of each other.
      (c) The OAS of systems with one NP planet and six QD satellites with different coupling $J_c$ between excitons.
      The dashed lines are calculated from Eq.~\eqref{eq:omegadc}.
      (d) The weight of both non-zero transition peaks extracted from (c).
      The parameters are set to $E_d\!=\!3.0eV$, $E_c\!=\!3.5eV$, and $\Delta_{dc}\!=\!80meV$.
    }
    \label{fig:6QDsJcW}
  \end{center}
\end{figure}

Figure~\ref{fig:6QDsJcW} shows the OAS of a single NP planet coupled to QD satellites.
This architecture should be easier to engineer in experiments and the number of QDs can be increased since the physical size of metallic NP is larger than QDs.
In general, the transition peaks and the corresponding frequencies can be determined analytically
\begin{equation}\label{eq:omegadc}
  \omega_{{d,c}}=\frac{1}{2}\left[ (\tilde{E}_d+\tilde{E}_c\pm\sqrt{(\tilde{E}_d-\tilde{E}_c)^2+4\tilde{\Delta}_{dc}^2} \right].
\end{equation}
where $\tilde{E}_{d(c)}\!=\!E_{d(c)}\!-\!2J_{d(c)}$, $\tilde{\Delta}_{dc}=\sqrt{N_\text{NP}N_\text{QD}}\Delta_{dc}$, and $N_{\text{QD}}(N_{\text{NP}})$ is number of QD(NP) in the system.
In the absence of $J_c$ and $J_d$, the OAS shows two transition peaks
$\omega_{{d,c}} \approx E_{d,c} \pm \frac{N_{\text{NP}}N_{\text{QD}}\Delta_{dc}^2}{E_d-E_c}$
when the system is far from resonance $(E_d-E_c)^2\!\gg\!4N_{\text{NP}}N_{\text{QD}}\Delta_{dc}^2$.
As the number of QDs in the system increases, shown in Fig.~\ref{fig:6QDsJcW}b, both peaks shift away from each other, which agrees with the above formula.
In Fig.~\ref{fig:6QDsJcW}c, the coupling between excitons is considered and the results show that there are still only two non-zero transition peaks.
Although there are more coupled QDs in the system, the uniform coupling between excitons ensures that only the zero-momentum state has non-zero coupling to plasmons.
Therefore, for isotropic couplings, the OAS from one zero-momentum excitation can always be determined from Eq.\eqref{eq:omegadc} by adjusting the energies and couplings accordingly where the details of this discussion are provided in the Methods section.
It turns out that the frequencies of both transition peaks change and the shifted amount can be determined from the energy of the zero-momentum state of the excitons.
More interestingly, the weight distribution between both peaks is affected by the couplings $J_c$ as well, shown in Fig.~\ref{fig:6QDsJcW}d.
Here, the weight of each transition is defined as the area beneath the peak and can be evaluated by integrating around the transition energy
\begin{equation}\label{eq:weight}
  \mathcal{W}(\omega_T)=\int_{\omega_T^-}^{\omega_T^+}d\omega\sigma(\omega).
\end{equation}
The transition amplitudes are normalized in the calculation so the total weight of all transition peaks in OAS is unity.
In the example shown in Fig.~\ref{fig:6QDsJcW}c, the weight shifts to the $\omega_c$ transition as the coupling $J_c$ increases.
However, the weight distribution depends on the difference of eigenenergies between plasmons and excitons, which are the diagonal elements in Eq.~\eqref{eq:omegadc}.
In this example, the zero-momentum state of excitons has eigenenergy $\tilde{E}_c\!=\!E_c\!-\!2J_c\!<\!E_c$.
As the $J_c$ increases, the difference of eigenenergies, $\vert\tilde{E}\!-\!E_d\vert$, becomes smaller and there will be more weight on the transition peak $\omega_c$.

\begin{figure}[th]
  \begin{center}
    \includegraphics[width=0.48\textwidth]{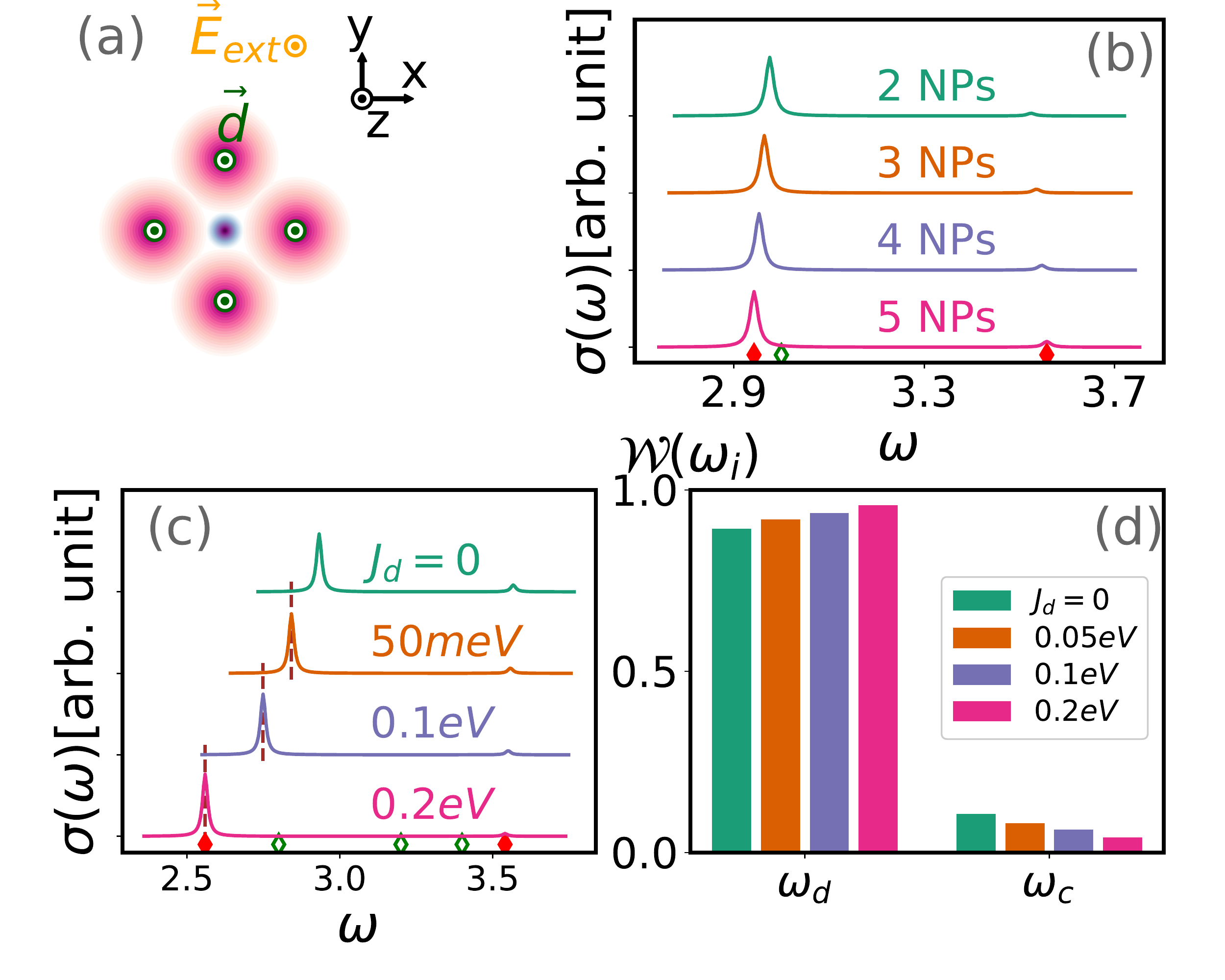}
    \caption{
      (a) The system has one QD planet and four NP satellites in the $x$-$y$ plane.
      The external probing field is on the $z$-axis.
      (b) The OAS of systems with one QD planet uniformly coupled to different numbers of NP satellites.
      The coupling between NPs are set to $J_d\!=\!0$.
      Both transition peaks $\omega_{d,c}$ shift with different numbers of NPs in the system, as given by Eq.~\eqref{eq:omegadc}.
      The empty (filled) diamond symbols under the curve mark the transition peaks (one excitation) with (non-)zero weight, and there are degenerate eigenvalues on top of each other.
      (c) The OAS of systems with one QD planet and six NP satellites with different coupling $J_d$ between plasmons.
      (d) The weight of both non-zero transition peaks extracted from (c).
      The parameters are set to $E_d\!=\!3.0eV$, $E_c\!=\!3.5eV$, and $\Delta_{dc}\!=\!80meV$.
    }
    \label{fig:6NPsJdW}
  \end{center}
\end{figure}

Next, we consider the QD-planet-NP-satellite structures.  (If the NP satellites were to form a complete metal cage, the external field would be completely screened, resulting in no polarization on the QDs.)
Although the screening may be strong, resulting in weak coupling between plasmons and excitons, the experiments show that the effect is still observable~\cite{Pein:2017fe,Karan:2015ft,Ji:2015bz}.
The results are shown in Fig.~\ref{fig:6NPsJdW} and again there are mainly two transition peaks as discussed previously - $\omega_{d}$, and $\omega_{c}$.
If the identity of the QDs and NPs are interchanged,
the frequencies of transition peaks are exactly identical in both structures under proper exchange of the couplings, but the transition weights will be different due to the initial photoexcited state.
The frequencies of both transition peaks shift as well if there are more NPs in the system and the dominant transition is still $\omega_d$.
In the presence of couplings between plasmons, the transition energy $\omega_d$ changes as shown in Fig.~\ref{fig:6NPsJdW}c.
Again, this problem can be treated by rotating the NP subsystem to new basis and determine the coupling to excitons accordingly.
For example, four NPs surround one QD illustrated in Fig.~\ref{fig:6NPsJdW}a, the eigenenergies of the NP subsystem are $E_d\pm 2J_d$ and $E_d$ (doubly degenerate) where only the eigenstate with eigenvalue $E_d\!-\!2J$ couples to the exciton with non zero matrix element $2\Delta_{dc}$.
It is worth mentioning that the eigenstate with eigenvalue $E_d\!-\!2J$ is also the zero-momentum state that the external photon excites.
Therefore, the system reduces to a simple problem with only one NP with energy $E_d\!-\!2J$ and coupling $2\Delta_{dc}$, and the transition frequencies can be determined from Eq.~\eqref{eq:omegadc} again.
These conclusions can be extended to many NPs as the couplings are uniform and only
the zero-momentum state is excited, so the energy shift is $E_d\!-\!2J$ and the coupling changes to $\sqrt{N_{\text{NP}}}\Delta$.
The detailed calculation and discussions are provided in the Methods section.
It is also important to look at the weight distribution of these two peaks.
In the shown example of Fig.~\ref{fig:6NPsJdW}c, as the coupling between plasmons $J_d$ increases, the weight of $\omega_d$ becomes more dominant, which is the opposite of increasing $J_c$ in the previous case.
The reason is the same though, from the difference of eigenenergies between plasmons and excitons.
Here, the energy of the zero-momentum state of plasmons has a lower energy $\tilde{E}_d\!=\!E_d\!-\!2J_d$ and a larger difference as $J_d$ increases, so the transition $\omega_d$ becomes more dominant.

\begin{figure}[th]
  \begin{center}
    \includegraphics[width=0.48\textwidth]{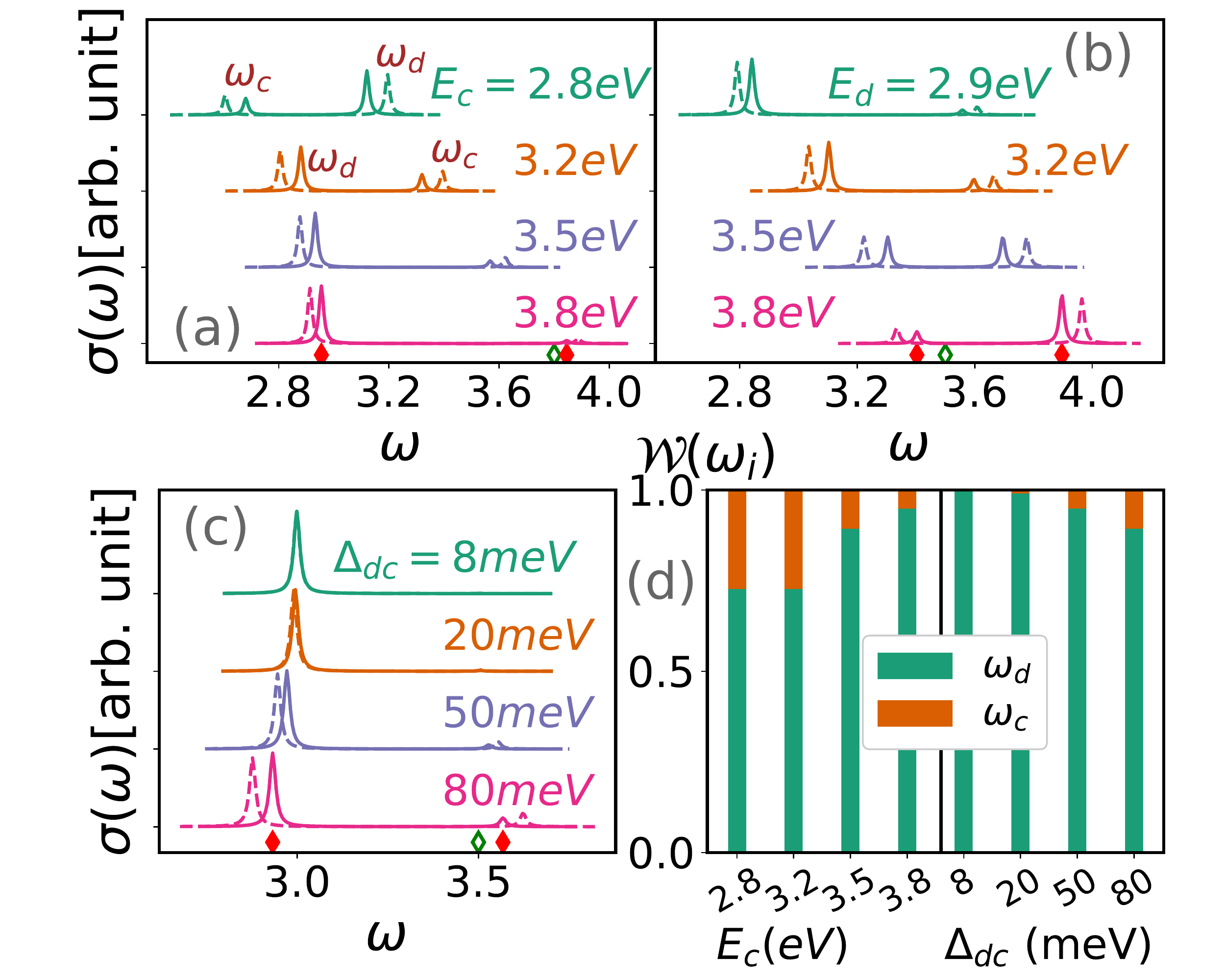}
    \caption{
      (a)-(c) The OAS of systems with six QDs surrounding one (solid line) and two (dashed line) NPs in the center with isotropic coupling between plasmons and excitons.
      The empty (filled) diamond symbols under the curve mark the transition peaks (one excitation) with (non-)zero weight, and there are degenerate eigenvalues on top of each other.
      (d) The weight of both transition peaks for three different $E_c$'s and $\Delta_{dc}$'s determined from (a) and (c).
      The parameters are set to $E_d\!=\!3.0eV$, $E_c\!=\!3.5eV$, $\Delta_{dc}\!=\!80meV$,
and $J_c\!=\!J_d\!=\!0$ unless indicated in the figures.
    }
    \label{fig:NP6QDsEcEdDcJc}
  \end{center}
\end{figure}

The intrinsic energy scales, like $E_d$, $E_c$, and $\Delta_{dc}$, also affect the spectrum as seen from Eq.~\eqref{eq:omegadc}.
The OAS of systems with six QDs surrounding one and two NP(s) are shown in Fig.~\ref{fig:NP6QDsEcEdDcJc} with varying intrinsic parameters.
The two nanoparticle case has the NP(s) on the z-axis, with one above and one below the xy-plane.
The two major transitions are located at $\omega_c$ and $\omega_{d}$.  Both peaks shift as $E_d$, $E_c$, or $\Delta_{dc}$ varied.
Moreover, as the energy of excitons approaches the energy of plasmons, the amplitude of the $\omega_c$ transition increases as shown in Fig.~\ref{fig:NP6QDsEcEdDcJc}a. The weight distribution of both peaks is shown in Fig.~\ref{fig:NP6QDsEcEdDcJc}d.
In the resonant case where $E_d\!=\!E_c$, the weight of both peaks are equal as shown in the $E_c\!=\!3.5eV$ curve in Fig.~\ref{fig:NP6QDsEcEdDcJc}b and the associated frequencies are $\omega_{c,d}\!=\!(E_d\!+\!E_c)/2\pm2\Delta_{dc}\sqrt{N_{\text{NP}}N_{\text{QD}}}$.
Here is another example showing that the distribution of the weight, which is determined from the overlap between eigenvectors, depends on the relative energy of plasmons and excitons, $\vert E_c\!-\!E_d\vert$.
This can also be seen in Fig.~\ref{fig:NP6QDsEcEdDcJc}d where the weight distributions of the transition peaks are the same for $E_c\!=\!3.4$ and $3.6eV$ when $E_d\!=\!3.5eV$.
In the resonant case, $E_c\!=\!E_d$, the weights of both peaks
are the same despite the fact that the shape and height can be different if the life time of plasmons and excitons differ.
More interestingly, if the coupling between plasmons and excitons $\Delta_{dc}$ is changed, the results clearly show that the weight of transition $\omega_c$ increases with the coupling strength as shown in Figs.~\ref{fig:NP6QDsEcEdDcJc}c and~\ref{fig:NP6QDsEcEdDcJc}d.
For systems with two NPs, the above statements are all still valid, and the only difference is the separation between two peaks which can still be captured by Eq.~\eqref{eq:omegadc}.
We also remark that the weight distribution from the coupling between plasmons and excitons can also occur in the QD-planet NP-satellite orientations.

\subsection*{General coupling}\label{sec:ang}
Here, we move on to more general cases and consider the effects from polarization of external electric field, the shape of NPs, and structures between QD(s) and NP(s) which should be able to capture the experiments.
First, all the NPs and QDs are in the $x$-$y$ plane and the external field is along the $x$-axis within the plane, so all the $z$ components on both NP and QD can be ignored.
Three different scenarios will be discussed here: QD-planet with needle like NP-satellite, NP-planet-QD-satellite, and QD-planet-NP-satellite.
In general, all the couplings are determined from the dipole-dipole interaction energy
\begin{equation}\label{eq:dd}
  U_{ij}=\frac{\hat{\vec{d}}_j\cdot\hat{\vec{d}}_i-3(\hat{\vec{d}}_j\cdot\hat{r})(\hat{\vec{d}}_i\cdot\hat{r})}{r^3}
\end{equation}
where $\hat{\vec{d}}_{j}$ is the dipole operator and $\vec{r}$ is the relative position vector between two dipoles.

\begin{figure}[th]
  \begin{center}
    \includegraphics[width=0.48\textwidth]{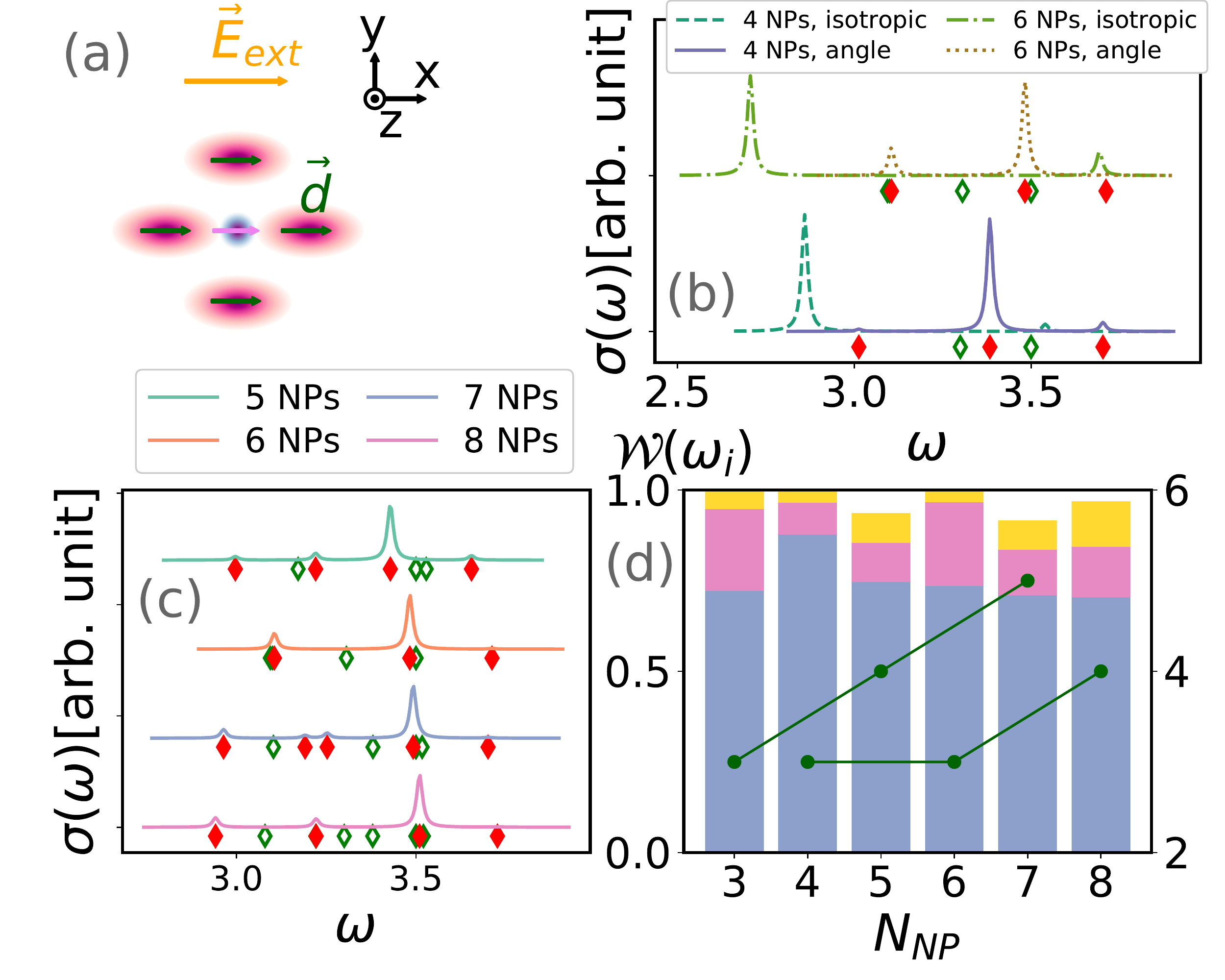}
    \caption{
      (a) The illustration of four needle-like NP satellites surrounding a single QD planet.
      Both couplings $J^{(ij)}_d$ and $\Delta^{(j)}_{dc}$ are angle dependent, given by Eq.~\eqref{eq:CoupleQDCenter}.
      (b) Comparison of OAS between isotropic and angle dependent couplings with four and six NPs.
      (c) The OAS of systems with different numbers of needle-like NP-satellites.
      The empty (filled) diamond symbols under the curve mark the transition peaks (one excitation) with (non-)zero weight, and there are degenerate eigenvalues on top of each other.
      (d) The weight (bars) of the three most dominant transitions and number of non-zero weight transitions (green circles) versus the number of NPs in the system.
      Three different colors in the bar graph show the first (blue), second (purple), and third (yellow) most dominant transitions.
      The system parameters are set to $E_c\!=\!3.5eV$, $E_d\!=\!3.3eV$, and $\Delta_{dc}\!=\!80meV$, and $J_d\!=\!0.2eV$.
    }
    \label{fig:QDCenterAngle1}
  \end{center}
\end{figure}

\subsubsection*{QD-planet with needle-like NP-satellites}
First, the system of one QD planet at the origin with index $i\!=\!0$ surrounded by several needle-like NPs is considered, shown in Fig.~\ref{fig:QDCenterAngle1}a.
Since the NPs have needle-like geometry and the external probing field is also along $x$ direction, we can consider that the only active mode on the NPs is $m\!=\!x$-component dipole moment~\cite{Artuso:2012vw}.
We also assume that all NPs are arranged on a ring with equal distance from the center of neighboring needles, thus the polarization of QD is mostly affected by the NP on $\phi\!=\!0,\pi$ if there are even number of NPs.
(Although we compute the effect of all NPs).
Therefore, it is reasonable to assume that the expectation value of the dipole moment on each NP to be in the
$\hat{x}$ direction as well as the one on the center QD.
The angle dependent coupling between nearest neighbor of plasmons and between plasmon and exciton are given by
\begin{eqnarray}\label{eq:CoupleQDCenter}
  \Delta_{dc}^{(j)} &=& -\Delta_{dc}P_2(\cos\phi_j),\\
  J^{(ij)}_d &=& -J_dP_2( \frac{\cos\phi_i-\cos\phi_j}{\sqrt{2(1-\cos(\phi_i-\phi_j))}} ),
\end{eqnarray}
where $P_2(x)=(3x^2-1)/2$ is second Legendre polynomial and $\phi_j$ is the azimuthal angle of NP-$j$.

The main results are summarized in Fig.~\ref{fig:QDCenterAngle1} for this scenario.
The comparison between isotropic and angle dependent couplings of this setup is also shown in Fig.~\ref{fig:QDCenterAngle1}b.
The results clearly show the qualitative difference between isotropic and angle dependent coupling.
For isotropic coupling, there are only two transitions with non-zero weight since only the zero-momentum state of the plasmons couple to the exciton.
The anisotropic coupling between plasmons
breaks the rotational symmetry of the ring geometry, so the momentum is not a good quantum number in general.
In addition to that, the anisotropic coupling between excitons and plasmons can make the problem even more complicated.
Although most of the spectral weight is distributed among the two dominant transitions, the frequencies of the two major transitions are shifted dramatically.
When there are more NPs in the system, shown in Fig.~\ref{fig:QDCenterAngle1}c, the weight distribution changes, so there are more visible peaks in the spectrum.
The number of non-zero weight transition peaks and the weight distributions are plotted in Fig.~\ref{fig:QDCenterAngle1}d.
With a reasonable number of NPs,
the number of non-zero weight transition peaks increases if there are an odd number of NPs, and the weight distribution changes when there are more NPs in the system.

\subsubsection*{NP-planet-QD-satellites}
\begin{figure}[th]
  \begin{center}
    \includegraphics[width=0.48\textwidth]{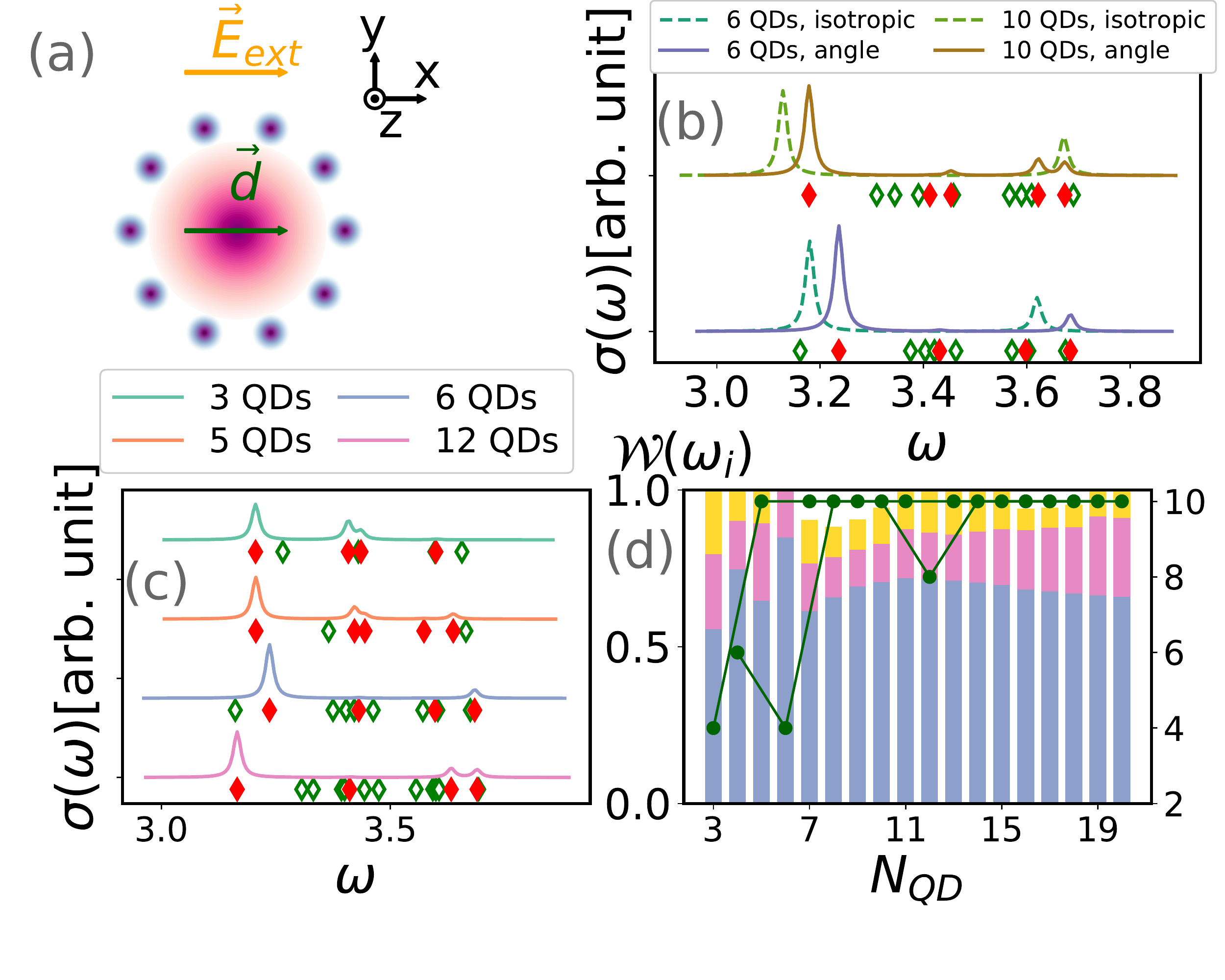}
    \caption{
      (a) Illustration of a system with ten QDs surrounding one NP.
      Both $J^{(ij)}_{c,mn}$ and $\Delta^{(i)}_{dc,mn}$ are angle dependent with two active modes.
      (b) The comparison of OAS between isotropic and angle dependent coupling with six and ten QD satellites.
      (c) The OAS of systems with different numbers of QD satellites.
      The empty (filled) diamond symbols under the curve mark the transition peaks (one excitation) with (non-)zero weight, and there are degenerate eigenvalues on top of each other.
      (d) The weights (bars) of the three most dominant transitions and the number of non-zero weight transitions (green circles) versus the number of QDs in the system.
      The three different colors in the bar graph show the first (blue), second (purple), and third (yellow) most dominant transitions.
      The parameters are set to $E_c\!=\!3.5eV$, $E_d\!=\!3.3eV$, $\Delta_{dc}\!=\!80meV$, and $J_c\!=\!0.1eV$.
    }
    \label{fig:NPCenterAngle}
  \end{center}
\end{figure}

Next, we consider only one NP planet at the origin with index $j\!=\!0$ and QD satellites as shown in Fig.~\ref{fig:NPCenterAngle}a.
All the QDs are the same distance from the NP, and spaced uniformly.
The external photon will create one excitation on the NP,
exciting the $x$-component of the dipole moment.
The NP and all QDs are spherically symmetric and the external field is in the plane.
The plasmons and excitons have two modes, $x$ and $y$-component dipole moments, and the couplings are all angle dependent and determined from the dipole-dipole interaction energy.
(The $z$-component is not excited in this geometry.)
A type of rotational symmetry is restored by including the two dipole components.
All the angle dependent couplings can be worked out.
The details are provided in the Methods section.

In Fig.~\ref{fig:NPCenterAngle}b, the results show the qualitative difference between isotropic and angle dependent coupling.
The number of transition peaks is no longer two if there are more than two QDs in the system.
For two QDs, the problem reduces to the uniform coupling case discussed previously.
The number of non-zero weight transition peaks is summarized in Fig.~\ref{fig:NPCenterAngle}d with up to twenty QDs.
Although there are more allowed transition peaks, most of the weight is still distributed among two major transitions.
The frequencies corresponding to the two major transitions also shift compared to isotropic coupling.
The frequencies of the transition peaks are determined by the eigenvalues of the exciton eigenstates with non-zero coupling to plasmons.
In Fig.~\ref{fig:NPCenterAngle}c, we present the OAS with different numbers of QD satellites.
For three QDs, the frequency of the major transition is still around $E_d$, but it shifts when there are more QDs in the system.
The weight distribution and number of non-zero weight transitions for up to twenty QDs are shown in Fig.~\ref{fig:NPCenterAngle}d.
For a reasonable number of QDs, there are ten non-zero transition peaks with most of the weight distributed among three dominant frequencies.
It is worth pointing out that the system with four or six QDs has fewer non-zero transitions, and it is possible that some special symmetry may apply to these cases.
The increasing number of non-zero weight peaks and the weight distribution are unique features of anisotropic coupling.

\begin{figure}[t]
  \begin{center}
    \includegraphics[width=0.48\textwidth]{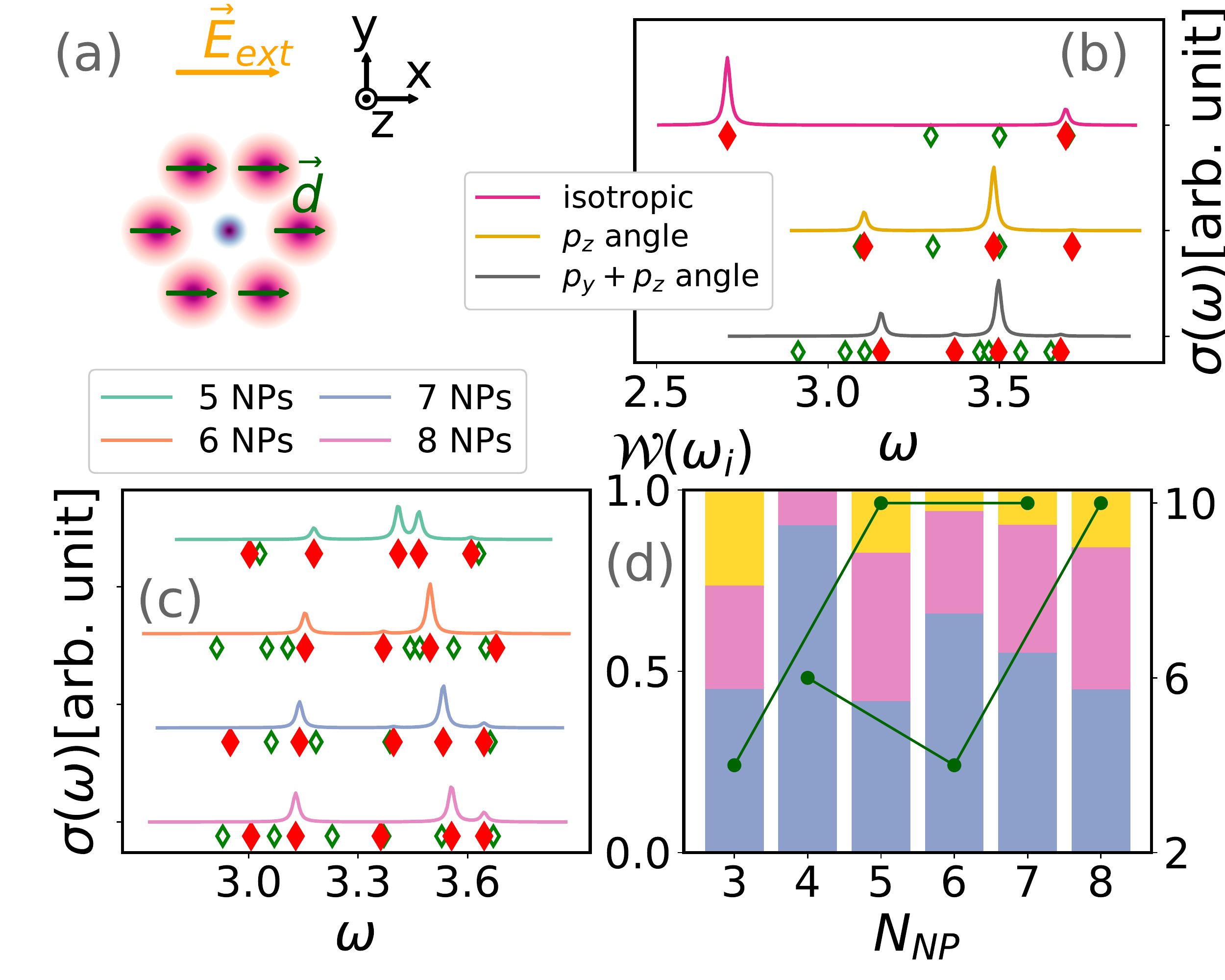}
    \caption{
      (a) Illustration of a system with six NPs surrounding a single QD.
      The green arrow indicates the zero momentum state excited by an external photon.
      (b) Comparison of OAS between isotropic, single mode, and two-mode angle dependent coupling with six NPs.
      (c) The OAS of systems with different numbers of NPs and two-mode angle dependent coupling.
      The empty (filled) diamond symbols under the curve mark the transition peaks (one excitation) with (non-)zero weight, and there are degenerate eigenvalues on top of each other.
      (d) The weight (bars) of the three most dominant transitions and number of non-zero weight transitions (green circles) versus number of NPs in the system.
      Three different colors in the bar graph show the first (blue), second (purple), and third (yellow) most dominant transitions.
      The system parameters are set to $E_c\!=\!3.5eV$, $E_d\!=\!3.3eV$, $\Delta_{dc}\!=\!80meV$, and $J_d\!=\!0.2eV$.
    }
    \label{fig:QDCenterAngle2}
  \end{center}
\end{figure}

\subsubsection*{QD-planet-NP-satellites}
The final scenario we consider is a single QD surrounded by spherical NPs, as shown in Fig.~\ref{fig:QDCenterAngle2}a.
In Fig.\ref{fig:QDCenterAngle2}b, the OAS of systems with six NPs under isotropic, single mode (needle-like NP), and two-mode angle dependent couplings are compared.
The two-mode model has a spectrum similar to the single mode one.
There is a small energy shift of both major transition peaks compared to the single mode model.
There are some transitions that are not present in the single mode model, as can be seen in the filled diamond symbols.
The major transition peaks have a small energy shift and different weight distribution, shown in Figs.~\ref{fig:QDCenterAngle2}c and~\ref{fig:QDCenterAngle2}d.
Also, the number of non-zero weight transition peaks is different from the single mode model when the number of NP satellites is the same.

\begin{figure}[th]
  \begin{center}
    \includegraphics[width=0.4\textwidth]{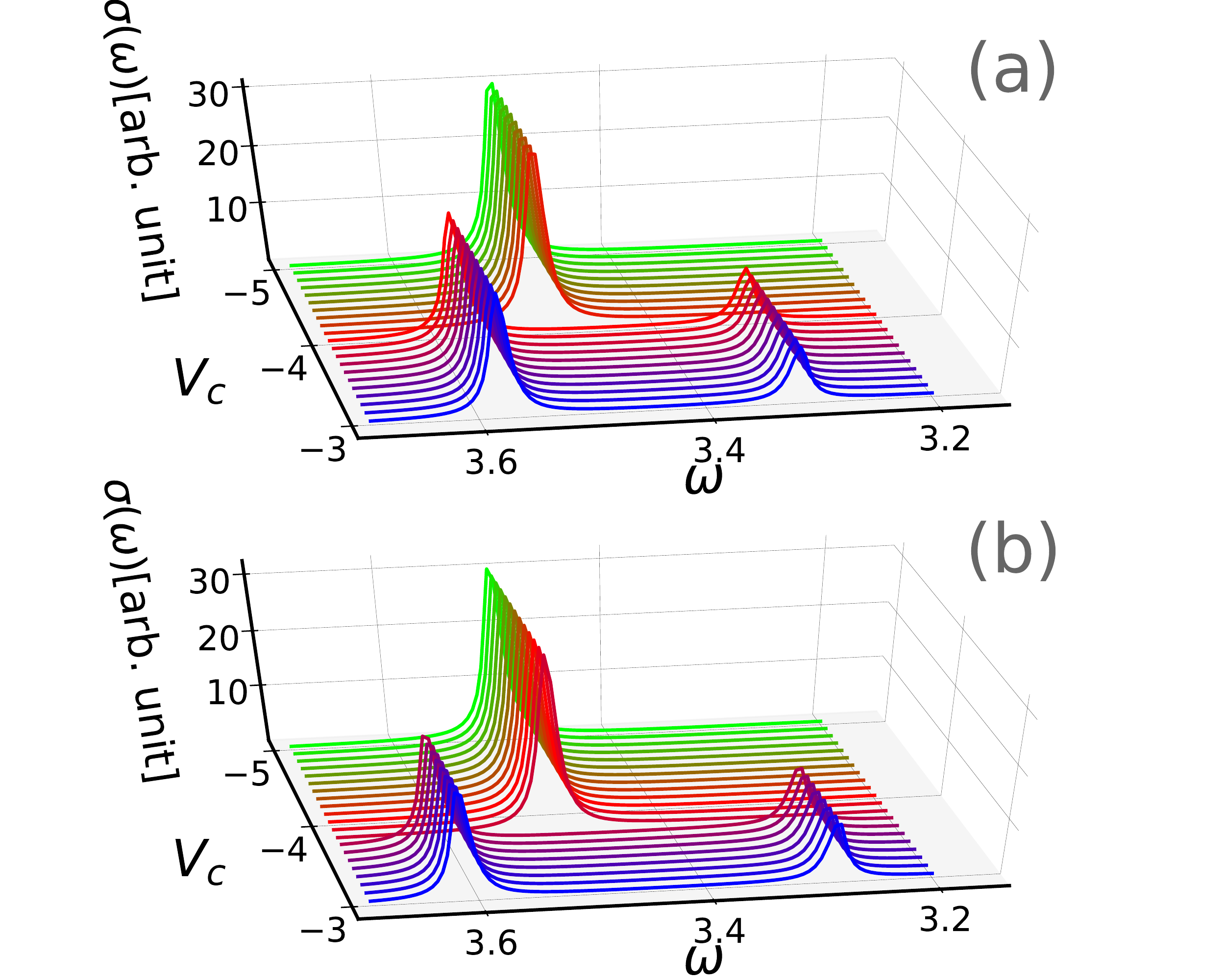}
    \caption{
      (a) Six and (b) ten QDs surrounding one NP with uniform coupling and interaction between excitons.
      The energy scales of plasmons $E_d\!=\!3.5eV$, excitons $E_c\!=\!3.4eV$, and $J_c\!=\!0$ are used.
    }
    \label{fig:6AttractionQDs}
  \end{center}
\end{figure}

\subsection*{Interaction between excitons}\label{sec:int}
In this final section, we discuss the effect of interaction between excitons on nearest neighbor QDs which is described by the Hamiltonian
\begin{equation}
  H_{\text{int}} = V_c\sum_{i}c^\dagger_i c_i c^\dagger_{i+1} c_{i+1},
\end{equation}
where the interaction $V_c$ can be either repulsive or attractive depending on the polarization of neighboring QD satellites.
The system Hamiltonian is now given by $H\!=\!H_{\text{NP}}\!+\!H_\text{QD}\!+\!H_{\text{int}}\!+\!H_{\text{dc}}$.
Here, we only consider the NP-planet-QD-satellite system and
a uniform interaction between nearest neighbor excitons.
If the interaction is repulsive, this additional energy scale does not affect the OAS.
This is because the OAS is determined by the vacuum state with no excitons and the one exciton sector,
and at least two excitons are required for $H_{int}$ to be nonzero.

On the other hand, an attractive interaction between excitons can change both the ground state and low energy excited states of the system.
Therefore, system with more than one excitation needs to be considered here.
In a $N$-QD system, for a sufficiently large attractive $V_c$, the initial state can be in the $N$ excitation sector and the OAS is determined from both the $N-1$ and $N+1$ excitation sector.
Figure~\ref{fig:6AttractionQDs} shows that there is a critical attraction that triggers the quantum phase transition which can be easily determined from ground state wavefunction.
Before the interaction crosses the critical value, the features of OAS do not change qualitatively.
As the interaction crosses the critical value, the ground state of the system is shifted from the vacuum to the state filled with excitons.
Thus, the feature coming from the coupling between plasmons and excitons is likely washed out.
For example, with six QDs surrounding one NP, the OAS changes dramatically when the attraction
reaches $-4.1eV$.
For the single NP cases, both $\omega_{d}$ and $\omega_c$ transitions merge into one with frequency equal to $E_d$, shown in Fig.~\ref{fig:6AttractionQDs}.
If there are ten QDs in the system, the critical value increases as shown in Fig.~\ref{fig:6AttractionQDs}b.
This indicates that the critical value depends on the number of QDs.
Similar findings are also found in the two NPs results, or when the plasmon-exciton couplings are angle dependent, when the interaction crosses a certain critical value.

\subsection*{Different absorption ratios between NPs and QDs}\label{sec:ratio}
\begin{figure}[t]
  \begin{center}
    \includegraphics[width=\columnwidth]{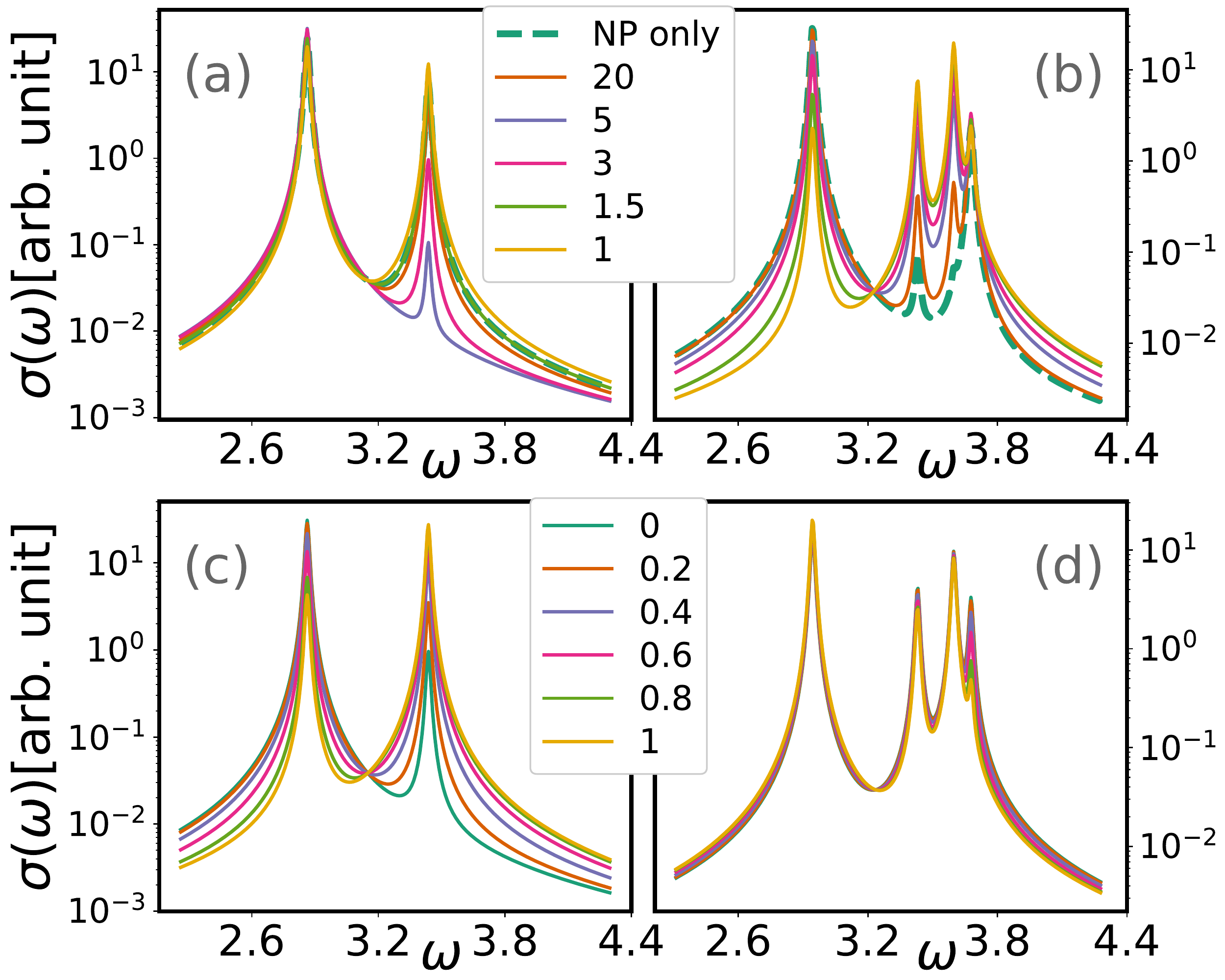}
    \caption{
      The OAS calculated with different absorption ratio between QDs and NP for (a,c) isotropic coupling with NP at the center and (b,d) general coupling with NP-planet-QD-satellites.
      (a)-(b) The ratio in the legend is defined as $C_{NP}/C_{QD}$ with $\beta\!=\!0$.
      (c)-(d) The dependency of the the relative phase $\beta/\pi$ with $C_{NP}/C_{QD}\!=\!3$.
      In all scenarios, there is one NP and six QDs with $E_d\!=\!3eV$, $E_c\!=\!3.5eV$, $\Delta_{dc}\!=\!0.1eV$, and $J_c\!=\!0.1eV$.
      The logarithmic scale is used on y-axis to emphasize the changes.
      }
    \label{fig:Ratio}
  \end{center}
\end{figure}

Here when  the probing photon field couples to both NP and QD simultaneously, we consider the absorption operator
$A\!=\!C_{NP}e^{i\beta}\sum_id_{im}+C_{QD}\sum_ic_{im}$, where $\beta$ is a relative phase shift.
It describes the excitation of $k\!=\!0$ state on both NPs and QDs where the ratio of absorption cross section depends on the size of the particles~\cite{leatherdale2002,yu2005,gaiduk2010,anderson2010}.
It is worth pointing out that the single excitation created by the external probing photon source will have an interference effect in the calculated spectrum.
This interference demonstrates that the dipole excited by the photon field on QD can affect the property of the NP through the coupling term, Eq.~\eqref{eq:NP-QDcoupling}, in the Hamiltonian.
Here, different absorption ratio is used to calculate the OAS for two scenarios: i) isotropic coupling with NP at the center and ii) general coupling with NP-planet-QD-satellites.
Although the locations of the peak will not change under this consideration, the weights of each peak will be affected by the different absorption rates.
Figure~\ref{fig:Ratio}(a) shows the effects on the absorption ratio for systems with isotropic coupling.
Comparing to the case that only NP gets excited by the external field, the signal from the QDs does not get amplified at the beginning as the ratio decreases, instead there are weights shifted toward $\omega_d$.
As the ratio keeps getting smaller and eventually reach equal absorption rate, the weights of $\omega_c$ get amplified and becomes larger than the one in the case where only NP is excited.
On the other hand, in the NP-planet-QD-satellites with general coupling shown in Fig.~\ref{fig:Ratio}(b), there are three significant peaks contributed from the presence of QDs and the weights of each peak change dramatically as the absorption rate decreases.
For only excited NP, the most dominant two are around $E_d$ and $E_c$, the weights from $\omega_d$ decreases along with the ratio and the the peak around $3.3eV$ get amplified.
By increasing the absorption rate of the QDs, the signals from QDs become stronger in the systems with general coupling.
Finally we remark that the influence of the QD dipole on the NP and the feedback of the NP dipole back to the QD, as modeled by an interaction between plasmons and excitons, can quantitatively change the optical absorption spectrum.

\section*{Conclusions}\label{sec:con}
In this article, a model of plasmon-exciton hybrid systems is proposed and its optical absorption spectrum is studied in detail.
Two different structures, with a NP or a QD as the planet, are considered.
Their spectra share similar features.
For isotropic coupling, a general recipe is given to calculate the frequencies of the two non-zero weight transitions.
The two allowed transitions are protected from the symmetry of the system and the excitation created by the external photon.
The weight distribution among the two transitions is also discussed as the intrinsic parameters are varied.
Interestingly, as the difference of eigenergies increases, the $\omega_d$ transition becomes more dominant.
The weight of the $\omega_c$ transition is maximum at $50\%$ of the total weight when the resonance condition $E_d\!=\!E_c$ occurs.
The coupling between the plasmon and exciton can change the weight distribution as well.

For angle dependent coupling, three different cases are discussed where all couplings are determined from the dipole-dipole interaction.
The spectra are qualitatively different than the isotropic case.
First, the number of non-zero weight transitions differ.
More transitions are allowed
since the rotational symmetry is broken by anisotropic coupling.
Secondly, the weight distribution changes when there are more satellite particles in the system.
The spectrum gets more complicated when there are more satellites in the system, in contrast to the isotropic coupling case where there are only two allowed transitions.
In addition, a phase transition is observed if an attractive interaction between excitons is present.
The critical point depends on the number of QDs in the system.
The spectrum changes when the ground state is lifted from the vacuum (zero excitation sector).
Finally, different absorption rate of NPs and QDs are considered in the isotropic and general coupling scenarios.
As the absorption rate changes, the weight distribution changes accordingly and the signals from QDs get amplified when the ratio besomes large enough in isotropic case and systems with general coupling.

In the article, we present the combined modeling of large scale hybrid systems and the ability to solve a fully quantum model.
The model captures the structures, geometries, and anisotropic features found in experiments.
The methods applied here also work for systems with less symmetry and even without any symmetry, including disorder, more than one excitation, or photoexcited states with mixed momentum.
By considering the anisotropic couplings between plasmons and excitons,
the OAS shows more features which can serve as a versatile tool to be a single photon source,
quantum information device, and  other functionalities of plasmon-exciton hybrid systems.

\section*{Methods}

\subsection*{Lanczos methods}\label{app:lanczos}
For small system sizes,
exact diagonalization can be used and
Eq.\eqref{eq:oas} can be solved by the sum over all states.
However, for large system sizes,
the Lanczos vector technique is more tractable~\cite{Hochbruck:2006cq,Moler:2003fn}.
The Hamiltonian can be approximated by
\begin{equation}
  H = \left(\begin{array}{ccccc}
  a_0 & b_0 & 0 & 0 & 0 \\
  b_0 & a_1 & 0 & 0 & 0 \\
  0 & 0 & \ddots & \ddots & 0 \\
  0 & 0 & \ddots & \ddots & b_{n-1} \\
  0 & 0 & 0 & b_{n-1} & a_n
  \end{array}\right)
\end{equation}
which is a tridiagonal matrix with matrix elements
\begin{eqnarray}
  a_n &=& \langle f_n \vert H \vert f_n \rangle, \\
  b_n &=& \vert\vert r_n \vert\vert.
\end{eqnarray}
Here, $\vert f_n\rangle$ are Krylov vectors which can be determined recursively by an input vector $\vert f_0 \rangle $
\begin{eqnarray}
  \vert r_n \rangle &=& (H - a_n)\vert f_n \rangle - b_{n-1} \vert f_{n-1} \rangle, \\
  \vert f_{n+1} \rangle &=& \frac{ \vert r_n \rangle }{ b_n }.
\end{eqnarray}
In order to capture all the transitions in the spectrum, we choose the input vector to be
$\vert f_0\rangle\!=\!\frac{A^\dagger \vert0\rangle }{ \vert\vert A^\dagger \vert0\rangle \vert\vert }$.
The tridiagonal matrix can be easily diagonalized and the identity can be approximated as
$I \approx \sum_n \vert \Phi_n \rangle\langle \Phi_n \vert$ where $\vert\Phi_n\rangle$ are the eigenstates of $H$.
The maximum $n$ indicates how many Lanczos vectors are used.
Inserting this identity into the spectrum of Eq.\eqref{eq:oas},
\begin{eqnarray}
  \sigma(\omega)
  &=& \frac{1}{\pi} \lim_{\eta\rightarrow0^+}
  \text{Im}\left\{
  \sum_{m,n}\langle 0 \vert A \vert \Phi_m \rangle \right.\\&&\left. \langle \Phi_m \vert \frac{1}{H- E_0 -\hbar\omega - i\eta}
    \vert \Phi_n \rangle\langle \Phi_n \vert A^\dagger \vert 0 \rangle  \right\} \nonumber\\
  &=& \frac{\langle0\vert AA^\dagger \vert0\rangle}{\pi} \\&& \lim_{\eta\rightarrow0^+}
  \text{Im}\left\{
  \sum_{n}\langle f_0 \vert \Phi_n \rangle \frac{1}{E_n - E_0 -\hbar\omega - i\eta} \langle \Phi_n \vert f_0 \rangle \right\} \nonumber\\
  &=& \frac{\langle AA^\dagger \rangle_0}{\pi} \lim_{\eta\rightarrow0^+}
  \sum_{n} \frac{\eta\vert \langle \Phi_n \vert f_0 \rangle \vert^2}{(\hbar\omega + E_0 - E_n)^2 + \eta^2} \\
  &=& \frac{\langle AA^\dagger \rangle_0}{\pi}
  \sum_{n} \delta(\hbar\omega + E_0 - E_n) \vert \langle \Phi_n \vert f_0 \rangle \vert^2 .
\end{eqnarray}
The above formula gives transition peaks at $\omega\!=\!E_n - E_0$, so the number of peaks are limited by number of Krylov vectors used, but do not exceed the dimension of the total Hilbert space of the system.
Also the weight of the transition will be determined by $\vert \langle \Phi_n \vert f_0 \rangle \vert^2$ with normalization factor $ \frac{\langle AA^\dagger \rangle_0}{\pi}$.
The summation of weight from all peaks will be unity, which also serves as a self check in this approach.

\subsection*{Uniform coupling systems}\label{app:uniform}
\begin{figure}[t]
  \begin{center}
    \includegraphics[width=\columnwidth]{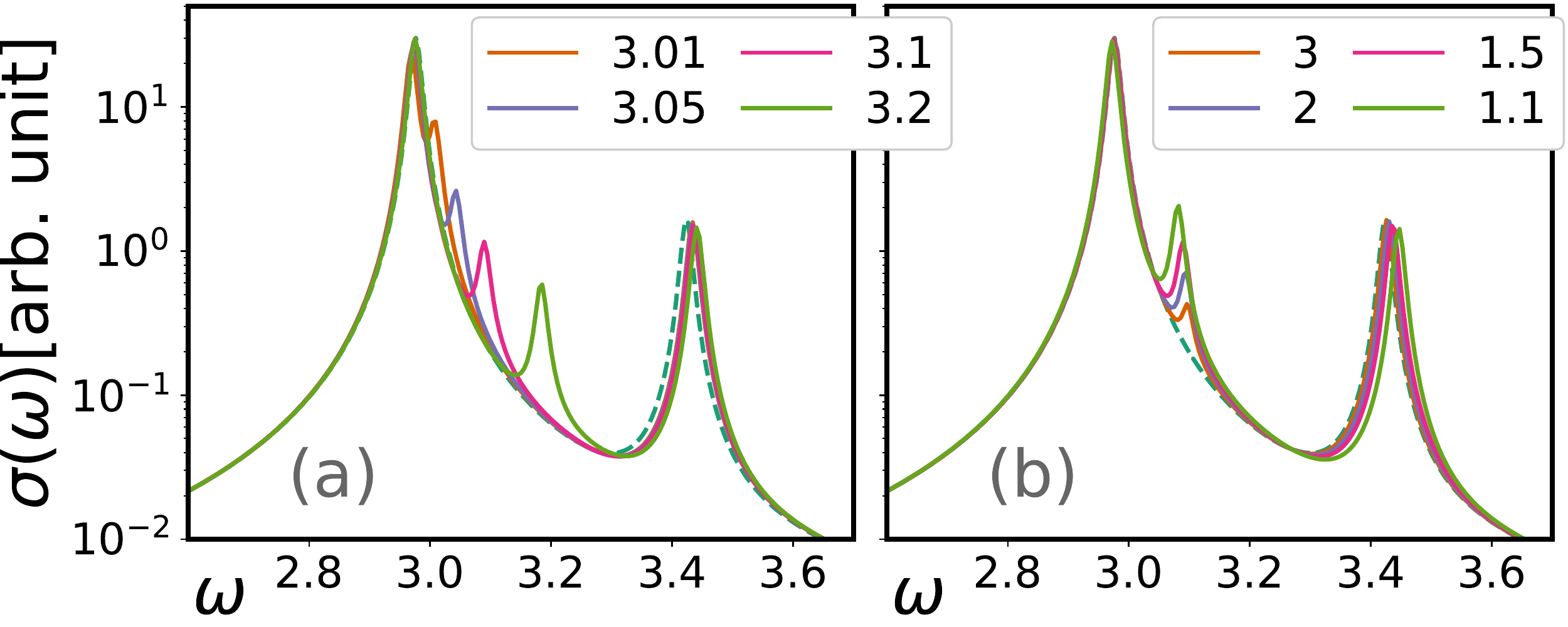}
    \caption{
      The OAS calculated with inclusion of quadrupole moment in the NP and one QD on the $z$-axis with isotropic coupling.
      Dashed line is the spectrum calculated without quadrupole plasmon.
      (a) $\Delta_{dc}\!=\!1.5\Delta_{qc}$ with different $E_{q}$'s compared.
      (b) $E_q\!=\!3.1eV$ with different $\Delta_{dc}/\Delta_{qc}$ ratios compared.
      Other parameters are set to $E_d\!=\!3eV$,  $E_c\!=\!3.4eV$, $\Delta_{dc}\!=\!0.1eV$, and $J_c\!=\!0$.
      The logarithmic scale is used on $y$-axis to emphasize the changes.
    }
    \label{fig:Quadrupole}
  \end{center}
\end{figure}

Here, we discuss the systems with uniform couplings in the one excitation sector.
The experimental setup will determine the spectrum.
In the theory, the non-zero matrix elements cause the transition peaks to appear.
Assuming the ground state is the vacuum state, the external photon only excites the zero-momentum state, and only one excitation is created in linear response.  One can consider
the system with only one particle,
either plasmon or exciton.
In general, the Hamiltonian of many NPs and many QDs is
\begin{equation}
  H=\left(\begin{array}{cc}
    h_{\text{NP}} & h_{\text{dc}} \\
    h^ \dagger _{\text{dc}} & h_{\text{QD}}
  \end{array}\right),
\end{equation}
where
\begin{equation}
  h_{\text{NP}}=\left(\begin{array}{ccccc}
    E_d & -J_d & 0 & \cdots & -J_d \\
    -J_d & E_d & -J_d & 0 & \cdots \\
    0 & \ddots & \ddots & \ddots & 0 \\
    \vdots & 0 & -J_d & E_d & -J_d \\
    -J_d & 0 & \cdots & -J_d & E_d
  \end{array}\right),
\end{equation}
with periodic boundary conditions.
For $h_{\text{QD}}$, the matrix is the same with parameters $E_c$ and $J_c$.
As for $h_{\text{dc}}$, it is a matrix with size $N_{\text{NP}}\!\times\!N_{\text{QD}}$ filled with elements $\Delta_{dc}$ everywhere.
First, we focus on the blocks of NPs ($h_\text{NP}$) and QDs ($h_\text{QD}$).
Translational invariance ensures that momentum is a good quantum number.  Thus one can rotate the basis to momentum space with eigenvalues $E_n\!=  \!E_d-2J_d\cos\frac{2n\pi}{N}$ with $n=0,1,\cdots,N/2$.
The eigenstates have momentum quantum number $k\!=\!2n\pi/N$.
Only the zero-momentum eigenstate of plasmons is excited.
Only the $k\!=\!0$ (zero-momentum) state of excitons has non zero matrix elements from $H_\text{dc}$ with amplitudes $-\Delta_{dc}\sqrt{N_\text{NP}N_\text{QD}}$ since the coupling between them do not alter the momentum quantum number.
Therefore, the problem reduces to a $2\!\times\!2$ eigenvalue problem with the matrix
\begin{equation}
  \tilde{H}=\left(\begin{array}{cc}
    \tilde{E}_d & -\tilde{\Delta}_{dc} \\
    -\tilde{\Delta}_{dc} & \tilde{E}_c
  \end{array}\right),
\end{equation}
where $\tilde{E}_{d(c)}\!=\!E_{d(c)}\!-\!2J_{d(c)}$ and $\tilde{\Delta}_{dc}=\sqrt{N_\text{NP}N_\text{QD}}\Delta_{dc}$.
This is identical to the problem of a system with one NP and one QD, and the eigenvalues are shown in Eq.~\ref{eq:omegadc}.
The transition weights can be also determined analytically from the corresponding eigenvectors, which is not elaborated here.
It is worth mentioning that this result is not valid once the symmetry is broken, including disorder in the system.
Here, one can also consider different dephasing time scales for the QDs and NPs by adding an imaginary part to the intrinsic energy level $E_{d(c)}\!\rightarrow\!E_{d(c)}-i\eta_{d(c)}$.
This implementation will change the resulting eigenvector.
Therefore, the amplitudes and shapes of the corresponding peaks will be different from each other.
Although we only consider the dipole-dipole interaction in the article, higher order multipolar moments can also be included in the proposed scheme and solved in a quantum mechanical manner.
For simplicity, we consider the quardrupole moment in the NP located at the origin with one quantum dot on the $z$-axis and the external probing field polarized in $z$-direction, thus only $l\!=\!2$ and $m\!=\!0$ components have non-zero matrix elements due to symmetry.
In this case, the Hamiltonian of NP is
\begin{eqnarray}
  H_{\text{NP}}=E_dd^\dagger d + E_qq^\dagger q
\end{eqnarray}
where $q^\dagger(q)$ the creation (annihilation) operator for quadrupole plasmons on NP.
The Hamiltonian of QDs is given in Eq.~\eqref{eq:Hqd}.
The couplings to QDs will involve both dipole and quadrupole moments on the NPs,
\begin{equation}
  H_\text{dc}=-\Delta_{dc}(d^\dagger c+h.c.) - \Delta_{qc}(q^\dagger c+h.c.).
\end{equation}
Figure~\ref{fig:Quadrupole} shows the comparison on the effects of quadrupole plasmons.
In short, the frequency around energy $E_{q}$ emerges and the effects from the quadrupole moment are not visible if the coupling between quadrupole plasmons and exciton ($\Delta_{qc}$) is small.

\subsection*{Angle dependent coupling}\label{app:angle}
The angle dependent couplings are determined from the dipole-dipole interaction energy in Eq.~\eqref{eq:dd}.
For the two-mode model, one needs to determine the coupling between neighboring satellites on the ring and between the planet and satellites for both modes.
We start from the couplings between the planet and satellites.
The operator nature of the dipole moments are left in the Hamiltonian and only the coupling forms are expressed here.
Assuming the QD is the planet, the $x$-$x$ couplings
\begin{eqnarray*}
  \Delta^{(j)}_{dc,xx}&=&\frac{1}{r_{dc}^3}\left[\vec{d}_j\cdot\vec{c}-3(\vec{d}_j\cdot\hat{r}_{d_jc})(\vec{c}\cdot\hat{r}_{d_jc})\right]\\
  &=&\frac{dc}{r_{dc}^3}(1-3\cos^2\phi_j) = \frac{-2dc}{r_{dc}^3}P_2(\cos\phi_j)\\
  &=&-\Delta_{dc}P_2(\cos\phi_j) ,
\end{eqnarray*}
where $\phi_j$ is the azimuthal angle of $j$-th satellite and the second Legendre polynomial is $P_2(x)=(3x^2-1)/2$.
The $y$-$y$ couplings have the same form but having $\sin\phi_j$ as Legendre variable.
For the $x$-$y$ modes, it is straightforward to arrive at
\begin{eqnarray*}
  \Delta^{(j)}_{dc,xy}=\Delta^{(j)}_{dc,yx}&=&\frac{1}{r_{dc}^3}\left[\vec{d}_j\cdot\vec{c}-3(\vec{d}_j\cdot\hat{r}_{d_jc})(\vec{c}\cdot\hat{r}_{d_jc})\right]\\
  &=&\frac{dc}{r_{dc}^3}(0-3\sin\phi_j\cos\phi_j) = \frac{-3dc}{r_{dc}^3}\sin\phi_j\cos\phi_j\\
  &=&-\frac{3}{2}\Delta_{dc}\sin\phi_j\cos\phi_j
\end{eqnarray*}
The couplings between neighboring satellites are
\begin{eqnarray*}
    J_{d,xx}^{(ij)}&=&\frac{1}{r^3_{dd}}\left[\vec{d}_i\cdot\vec{d}_j-3(\vec{d}_i\cdot\hat{r}_{ij})(\vec{d}_j\cdot\hat{r}_{ij})\right]\\
    &=&\frac{d^2}{r^3_{dd}}\left[1-3( \frac{\cos\phi_i-\cos\phi_j}{\sqrt{2-2\cos(\phi_i-\phi_j)}} )^2\right]\\
    &=&\frac{-2d^2}{r^3_{dd}}P_2( \frac{\cos\phi_i-\cos\phi_j}{\sqrt{2-2\cos(\phi_i-\phi_j)}} )\\
    &=&-J_dP_2( \frac{\cos\phi_i-\cos\phi_j}{\sqrt{2-2\cos(\phi_i-\phi_j)}} ),
\end{eqnarray*}
where the operator value of dipole moment is $\vec{d}\!=\!d\hat{x}$ and the position vector is $\vec{r}_{ij}\!=\!\vec{r}_i-\vec{r}_j\!=\!r(\sin\phi_i-\sin\phi_j)\hat{x}+r(\cos\phi_i-\cos\phi_j)\hat{y}$.
For the $y$ component $\vec{d}=d\hat{y}$, the couplings have the following form
\begin{eqnarray*}
    J_{d,yy}^{(ij)}&=&\frac{1}{r^3}\left[\vec{d}_i\cdot\vec{d}_j-3(\vec{d}_i\cdot\hat{r}_{ij})(\vec{d}_j\cdot\hat{r}_{ij})\right]\\
    &=&\frac{d^2}{r^3}\left[1-3( \frac{\sin\phi_i-\sin\phi_j}{\sqrt{2-2\cos(\phi_i-\phi_j)}} )^2\right]\\
    &=&\frac{-2d^2}{r^3}P_2( \frac{\sin\phi_i-\sin\phi_j}{\sqrt{2-2\cos(\phi_i-\phi_j)}} )\\
    &=&-J_dP_2( \frac{\sin\phi_i-\sin\phi_j}{\sqrt{2-2\cos(\phi_i-\phi_j)}} ).
\end{eqnarray*}
Finally, the coupling between $x$ and $y$ components on nearest neighboring satellites is also non-zero,
\begin{eqnarray*}
    J_{d,xy}^{(ij)}=J_{d,yx}^{(ij)}&=&\frac{-3d^2}{r^3}(\sin\phi_i-\sin\phi_j)(\cos\phi_i-\cos\phi_j)\\
    &=&-\frac{3}{2}J_d \frac{(\sin\phi_i-\sin\phi_j)(\cos\phi_i-\cos\phi_j)}{2(1-\cos(\phi_i-\phi_j))}.
\end{eqnarray*}
For the NP as planet, the above derivation holds as well.

\section*{Conflicts of interest}
There are no conflicts to declare.

\section*{Acknowledgements}
  We thank Dmitry Yarotski for fruitful discussions.
  This work is supported by the Center for Integrated Nanotechnologies, a U.S. DOE BES user facility, in partnership with the LANL Institutional Computing Program for computational resources.





\bibliography{LANL,rsc}

\begin{mcitethebibliography}{58}
\bibitem[not()]{note2}
An uniform external electric field interacting with a conducting sphere creates
  a dipole moment, but not higher multipole moments. For interactions between
  particles, the dipole-dipole interaction dominates over interactions from
  higher multipoles as the distance increases.\relax
\mciteBstWouldAddEndPunctfalse
\mciteSetBstMidEndSepPunct{\mcitedefaultmidpunct}
{}{\mcitedefaultseppunct}\relax
\EndOfBibitem
\end{mcitethebibliography}
\bibliographystyle{rsc} 

\end{document}